\documentclass[twocolumn]{aastex631}

\newcommand{\eb}{\begin{equation}}
\newcommand{\ee}{\end{equation}}
\newcommand{\masyr}{mas yr$^{-1}$}
\newcommand{\masyrtwo}{mas yr$^{-2}$}
\newcommand{\kms}{km~s$^{-1}$\,}
\newcommand{\msun}{$M_\odot$\,}
\newcommand{\uasyr}{$\mu$as yr$^{-1}$\,}

\definecolor{rkka}{RGB}{219,66,32}
\usepackage{amsmath}
\usepackage{epstopdf}

\shorttitle{Candidate stellar mass black holes in Hipparcos and Gaia}
\shortauthors{Makarov \& Tokovinin}

\begin{document}

\title{Verification of astrometrically accelerating stars from Hipparcos and Gaia: I.  Methodology and application to HIP 44842}

\correspondingauthor{Valeri V. Makarov}
\email{valeri.makarov@gmail.com}

\author[0000-0003-2336-7887]{Valeri V. Makarov}
\affiliation{U.S. Naval Observatory, 3450 Massachusetts Ave NW, Washington, DC 20392-5420, USA}
\author{Andrei Tokovinin}
\affiliation{Cerro Tololo Inter-American Observatory—NSF’s NOIRLab Casilla 603, La Serena, Chile}

\begin{abstract}
A large number of candidate binary stars with apparent acceleration on the sky has emerged from analysis of astrometric data collected by the Hipparcos, Tycho-2, and Gaia space missions. Although the apparent acceleration can serve as a relatively reliable indicator of binarity, it provides scarce information about the orbital and physical parameters of the components. With an emphasis on the search for stellar-mass black holes and neutron stars hidden in binary systems, we start a broader effort to characterize the most promising candidates using follow-up ground-based observations. Accurate quantification of orbital and physical parameters of systems with dim or invisible companions requires combination of Hipparcos, Gaia, and precision spectroscopic measurements. In this paper, we review the necessary steps in this implementation and describe the improved Hipparcos-Gaia sample of long-term astrometric accelerations which includes correction of sky-correlated systematic errors using the vector spherical decomposition method. As an example, we study one Hipparcos star with a large acceleration, HIP 44842, where the companion is revealed to be a normal main sequence star.
\end{abstract}

\section{Introduction} \label{section: Introduction}

Stellar mass black holes (StMBHs) are the end products of massive-star evolution.  Most massive stars belong to multiple systems, and StMBHs with close companions 
(orbital periods $\le 10$ d) are very conspicuous because of the high-energy emission in X-rays fueled by accretion from the donor star. Furthermore, presence of a massive companion can be asserted by radial velocity (RV) variation in tight pairs \citep{1992Natur.355..614C}. Existence of wider (periods $>30$ d) binaries  containing  non-accreting (dormant) StMBHs is expected on the general grounds, because the median period of massive binaries is much longer than 30 d. Such objects are detectable via the reflex motion of their optical companions revealed through astrometric accelerations and/or variable RVs. Asymmetric mass ejection associated with the BH formation imparts ``kicks'' that can disrupt wide binaries. The magnitude of these kicks is uncertain, and discoveries of relatively wide binaries containing dormant  StMBHs (DBHs) provide useful constraints on the natal kicks. Our work complements other recent efforts in this area. 

DBHs are rare, so a large number of targets should be screened to detect them.  Although \citet{2002ApJ...572..944G} estimated that that $30\,f_{\rm FSN}$ DBHs should be present among the astrometric binaries in Hipparcos, none have been reported. Here 
$f_{\rm FSN}$ is the probability of a massive star (between 8 and 30 \msun) to fail as a supernova \citep{1978Ap&SS..57..245M, 1993ApJ...405..273W}.  The conclusion was that  $f_{\rm FSN}$ must be very low.  All sky astrometric surveys, such as Hipparcos and now Gaia, provide new opportunities for this screening using the astrometric apparent acceleration parameter as the main indicator \citep{2003AN....324..419K}. Using this method, a dedicated search in the Gaia DR3 \citep{Gaia3} data resulted in the discovery of two DBH systems, confirmed by spectroscopic follow-up measurements \citep{2023MNRAS.518.1057E, 2023MNRAS.521.4323E}. \citet{2023AJ....165..193W} used proper motions from the two Gaia data releases, DR2 and DR3, and the epoch difference between them, to identify some 30 thousands probable accelerating systems in the Solar vicinity, extending the search beyond the Hipparcos sample.

Progenitors of StMBHs are massive ($>$8 \msun) stars. Their binary companions are also expected to be relatively massive, based on the available binary statistics. So,  targets for the search of DBHs are usually selected among massive stars: either young OB stars or products of their evolution --- red giants. Orbital periods of binaries composed of a giant and a StMBH are expected to be longer than $\sim$100 days, otherwise they would be shortened by the common envelope evolution. The orbital period of the first such known pair, Gaia BH2, is 1277 days \citep{2023MNRAS.521.4323E}. Only a small fraction of binaries containing StMBHs have  a  chance to survive the evolution and remain discoverable as DBH systems. \citet{2023MNRAS.518.1057E,2023MNRAS.521.4323E}  further clarify why only two DBH binaries have been identified so far among the $10^5$ astrometric binaries in Gaia DR3. 

Large input samples are needed to discover rare objects. However, the presence of rare outliers increases the chances of false positives caused by errors in the data. Such was the case of our previous attempt to find DBHs among $10^3$ red giants: the four candidates with large RV variation have not passed verification by additional spectroscopy; they were caused by the data glitches \citep{2019AJ....157..136M}. 

\citet{Mahy2022} searched for DBH companions to 32 Galactic OB stars known as single-lined spectroscopic binaries (SB1s) with periods in the range 2--55 days. They detected lines of stellar secondaries in 17 systems, converting them into SB2s, confirmed the known BH companion in the close binary Cyg X-1 (used as a test case), and found another strong DBH candidate, HD~130298, with an orbital period of 14.6 days. 

%with a similarly large mass of the ``dark'' secondary, 7 \msun.
% input sample from GOSS, clusters, Barda. Periods 1.9-23 days. HD 130298: 14.6 days
% Cite Heger2003, no mention of failed SNe
% X-quiet: MWC656 (Casares2014) and HD 96670 (Gomez, Grindlay 2021).
% predictions: 3% dormant BH (Langer 2020)
% Review Casares, Jonker 2014

\citet{Janssens2023} considered false positives encountered in the search for DBH based on the Gaia DR3 astrometry, specifically the blue supergiant companions and triple systems. A false-positive candidate containing a stripped core (product of the mass transfer) is presented by \citet{Zak2023}; this 34.54-day binary was originally identified as an ellipsoidal variable V1315 Cas, and the presence of a DBH companion has been suggested. Detection of a DBH in the 60-day binary MWC656 with a Be primary component has been contested by \citet{Janssens2023b}. These few examples illustrate that the search for DBH is currently a very active field. 

% Blue supergiant companions, triples. Current astro-orbits in DR3 have too stringent plx constraints,

The 3-year time span of the Gaia DR3 data restricts the periods of astrometric orbits available so far. Binaries with massive invisible companions and longer periods are detectable via combined analysis of the Gaia DR3 and Hipparcos data. We present such analysis below in Section~\ref{sele.sec}. A dozen of stars with large accelerations are mostly false positives, but we find a few promising candidates. In Section~\ref{section:HIP44842}, we test one candidate with a large acceleration signal, the red giant HIP~44842, by monitoring its RV, and show that its companion is a normal star rather than a StMBH. Our work is summarized in Section~\ref{sec:summary}.

\section{Selection of candidate accelerating stars}
\label{sele.sec}

Utilizing the principle of astrometric acceleration in unresolved binaries with fainter or dim companions, we start with the general sample of Hipparcos-Gaia stars that show acceleration signals in the available data. Although previous publications provide similar selections \citep{2019A&A...623A..72K, Brandt2021, 2022A&A...657A...7K}, we introduced significant improvements in this procedure hoping to achieve the highest degree of accuracy and reliability, as described in this Section. We begin with the main Hipparcos catalog \citep{1997A&A...323L..49P}, which includes 117955 entries. For each star, the mean epoch position $(\alpha, \delta)$ is propagated from 1991.25 to 2016.0 using the measured mean epoch proper motions, when available. It is essential to use the accurate vectorial epoch transformation routines for this analysis \citep{1983veas.book.....M, 2014A&A...570A..62B}, although we chose to omit the relativistic and light-time terms, which are quite small for the vast majority of the Hipparcos stars. Instead of making a one-way transfer of Hipparcos positions to 2016, we use a two-step transfer bringing the Hipparcos 1991.25 positions onto the ``midway" epoch 2000, and using the transformed Gaia DR3 \citep{Gaia3} positions on the same epoch. This allows us to minimize the possible loss of objects with extreme accelerations and with inaccurate proper motions in either catalog. An accommodating search radius of $2\arcsec$ was chosen for the same reason.

Out of 117,955 stars in Hipparcos with proper motions, 433 do not have any counterparts within $2\arcsec$ in Gaia DR3 at epoch 2000. We inspected this sample and determined that most of them are flagged binaries of type C (resolved double stars) and X (stochastic solutions), as well as very bright stars with perturbed results or missing data in Gaia DR3 (such as Betelgeuse and Antares). Many of these missing objects do not have proper motions or parallaxes in Gaia. HIP 1242 is a typical example. We find the true counterpart for this X-type binary in Gaia at a distance of $22.126\arcsec$ from the Hipparcos position, position angle $134\degr$. The difference in position implies a proper motion perturbation much greater than the threshold value $80.8$ \masyr\ over 24.75 yr implemented in our analysis. A perturbation of nearly 1 as yr$^{-1}$ for this nearby M dwarf is hard to explain in physical terms. Generally, the stochastic X-solutions in Hipparcos are believed to be binaries with shorter periods ($< 4$ yr) with complicating factors such as photometric variability, which require dedicated treatment and robust orbital fitting \citep{2006ApJS..166..341G, 2007ApJS..173..137G}. We decided to remove from this study  all flagged binaries of types C (resolved double stars with fixed positions), X (stochastic, or failed solutions), and O (systems with orbital solutions) with unreliable astrometric data, leaving 102,828 stars. Of these, 102,549 have proper motions in Gaia DR3.

Following the paradigm established in the Tycho-2 catalog \citep{2000A&A...357..367H}, we compute long-term proper motions using the mean positions in Hipparcos and Gaia separated by 24.75 yr. Specifically, the position difference vector $\Delta \boldsymbol{r}=\boldsymbol{r}_{\rm G}-\boldsymbol{r}_{\rm H}$ is projected on the local tangential basis vectors $\boldsymbol{\tau}_\alpha$ and $\boldsymbol{\tau}_\delta$ at the Gaia position and divided by 24.75. This derived proper motion vector is called $\boldsymbol{\mu}_{\rm HG}$ in formulae or HGPM in graphs. Its formal variance for each star is computed as the direct sum of the two position covariance matrices divided by 24.75$^2$. Accelerating stars emerge when these long-term proper motion vectors are compared with the nominal short-term proper motions in Gaia, called $\boldsymbol{\mu}_{\rm G}$ or GPM. The difference between these vectors $\Delta\boldsymbol{\mu}$ is the main acceleration signal, and its formal variance $\boldsymbol{c}_{\Delta\boldsymbol{\mu}}$ includes also the covariance of the Gaia proper motion.

\begin{figure*}
\includegraphics[width=0.47 \textwidth]{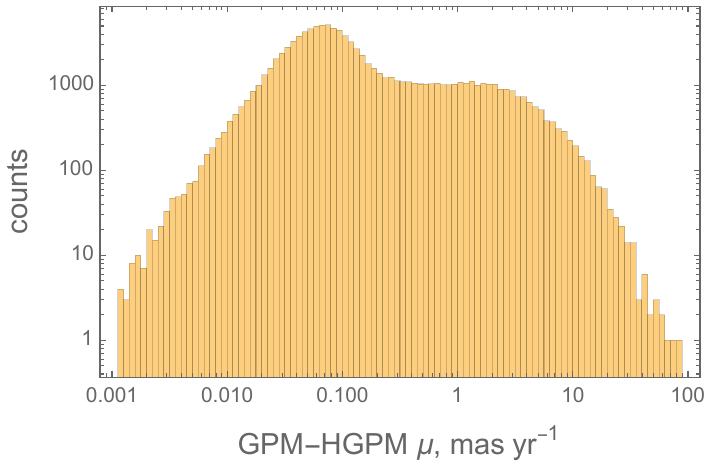}
\includegraphics[width=0.47 \textwidth]{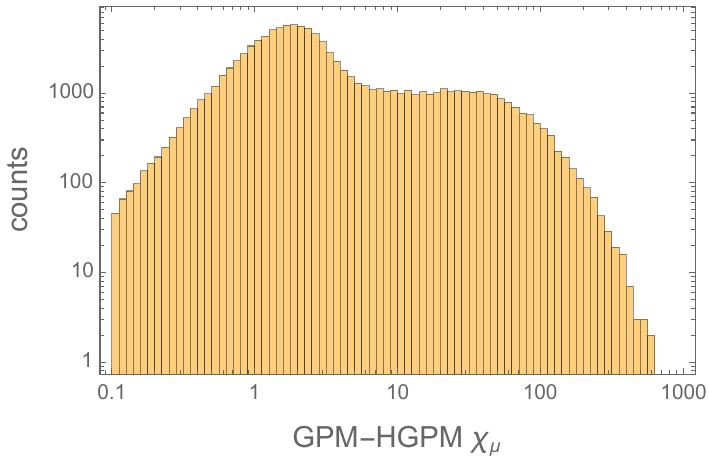}
\caption{Left panel: Histogram of GPM $-$ HGPM proper motion differences for 102,549 Hipparcos/Gaia stars in \masyr. Right
  panel: histogram of standardized GPM $-$ HGPM proper motion differences for the same sample of stars. Note the logarithmic scale of both axes.
\label{dmu.fig}
}
\end{figure*}

Fig. \ref{dmu.fig} shows the resulting histograms for two variables of interest, the magnitude of the $\Delta\boldsymbol{\mu}$ vector (left plot) and the standardized variance $\chi_{\boldsymbol{\mu}}$ (right plot), which is computed as
\eb
\chi_{\boldsymbol{\mu}}=\sqrt{\Delta\boldsymbol{\mu}^T\;\boldsymbol{c}_{\Delta\boldsymbol{\mu}}^{-1}\;\Delta\boldsymbol{\mu}}.
\ee  
This value is expected to be distributed as $\chi(2)$ for unperturbed stars with only random observational errors. Both sample distributions are abnormal. They show a prominent secondary hump at higher standardized and absolute values, which is accentuated by the logarithmic scale. This secondary population undoubtedly includes mostly components of binary and multiple systems with orbital acceleration on the time scales $<100$ yr. The mode of the $\chi_{\boldsymbol{\mu}}$ distribution is shifted from the expected 1 to 1.4. This shift can be explained in two different ways. Physically, the majority of stars may show small acceleration signals if they have hidden companions of planetary or brown dwarf mass. Technically, the formal measurement errors in the involved catalogs may be underestimated by approximately 40\%. For the latter option, we note that the main contributor to the formal uncertainty of HGPM is, statistically, the Hipparcos position. The degraded performance for the bulk of the sample is the critical limiting factor for a reliable and accurate identification of accelerating stars. Therefore, we have to take any available steps in refining the method and correcting possible astrometric error overheads.

\subsection{Objects with the largest magnitudes of acceleration}
\label{30.sec}
We find clear evidence of accelerating binary stars in the collection of HGPM proper motions in comparison with the short-term Gaia DR3 proper motions. Some stars stand out with very large signals. For example, 16 stars have tangential velocity differences, computed as
\eb
\delta_v=4.74\,\delta_\mu/\varpi,
\label{deltav.eq}
\ee
where $\delta_\mu=\|\Delta \boldsymbol{\mu}\|$, and $\varpi$ is the parallax, in excess of 30 km s$^{-1}$. These may be considered as prime candidates for accelerating binaries with invisible BH companions. However, most of these candidates are relatively distant stars with small parallaxes, where the $\delta_\mu$ signal is also small in absolute units. The computed $\chi_{\boldsymbol{\mu}}$ value may be deceptive, because a relatively small systematic astrometric error in the proper motions involved in this computation can perturb the result and bias the selection. Furthermore, an error in Gaia parallax, which is in the denominator of Eq. \ref{deltav.eq},  can artificially boost the estimated velocity difference for distant stars. The sky-correlated systematic errors of proper motions can be removed from the computed $\delta_\mu$ values for each star, as described in Section \ref{syst.sec}. This correction results in a smaller selection of 14 candidate systems listed in Table \ref{30.tab}. We provide additional notes about several most conspicuous candidates.

\begin{deluxetable}{c c llll }    
\tabletypesize{\scriptsize}     
\tablecaption{Candidates with velocity signals above 30 \kms\
\label{30.tab}          }
\tablewidth{0pt}                                   
\tablehead{                                                                     
\colhead{HIP} & 
\colhead{Gaia id} & 
\colhead{$\varpi$} & 
\colhead{$\sigma_\varpi$} & 
\colhead{$\delta_v$} & \colhead{$\chi_\mu$}\\
& & mas & mas&
\kms\ & 
}
\startdata
 5714 & 4687159868124113152 & 0.013 & 0.02 & 44.2 & 6.7 \\
 22365 & 204531088580182016 & 0.1 & 0.041 & 35.4 & 145.8 \\
 27655 & 4657362548954664192 & 0.012 & 0.022 & 97.3 & 11.7 \\
 37458 & 5292451705080043776 & 0.735 & 0.476 & 38.8 & 122.8 \\
 65385 & 6068221514276625408 & 0.329 & 0.445 & 34.4 & 47.3 \\
 73727 & 5887671362760158080 & 0.239 & 0.425 & 37.4 & 12.5 \\
 74063 & 6255787649521569408 & 0.419 & 0.083 & 31.7 & 9.1 \\
 78765 & 5996831633044142208 & 0.267 & 0.489 & 60.4 & 41.7 \\
 88838 & 4069737050648231424 & 0.05 & 0.187 & 165.5 & 77.9 \\
 89018 & 4526441470342242304 & 0.097 & 0.064 & 33.2 & 5.1 \\
 89440 & 4094031138050641408 & 0.064 & 0.324 & 86. & 22.9 \\
 94706 & 4087778009231713408 & 0.255 & 0.15 & 61. & 257.3 \\
 97610 & 6864873901123780736 & 0.624 & 0.557 & 37.5 & 170.5 \\
 113665 & 1934952590134811392 & 0.726 & 0.417 & 34.5 & 237.2 \\
\enddata 
\end{deluxetable}

HIP 5714 = HD 7583 is an A0I type supergiant and the brightest star in the Small Magellanic Cloud. It is chemically peculiar showing a deficit of metals in the spectrum \citep{1972MNRAS.159..155P}. Epoch photometry is available in Gaia DR3 for this star, which shows small variations in $G$ between 10.075 and 10.152 mag. The closest companion on the sky separated by $4\arcsec$ is marginally faint ($G=20.48$ mag), so it cannot be a threat to Gaia astrometry. The Gaia parallax value ($\varpi=0.012$ mas) is consistent with the distance to the SMC within the formal error. An upward correction of the parallax can push the estimated $\delta_v$ below 30 \kms, but it still remains high.

HIP 22365 = KS Per = HD 30353 is a hydrogen-deficient high-velocity mover and a spectroscopic binary with a large radial velocity semiamplitude $K$ \citep{1950ApJ...111..333B, 1988A&AS...75..157M}. It is also a supergiant of type A. The small Gaia-determined parallax $\varpi=0.100\pm 0.041$ mas combined with substantial proper motion indicates a high space velocity in excess of 100 \kms. The spectrum, beside very weak hydrogen lines, shows strong lines of ionized metals and radial velocity variation with a period $\sim 1$ yr and $K> 50$ \kms. The first set of orbital elements was derived by \citet{1955AJ.....60..162H}. \citet{1982A&A...113L..22D} found no trace of the companion in the optical spectrum but detected its presence in the far-UV. They suggested that the primary is a helium giant of about $1\,M_{\sun}$ in or near its Roche lobe overflow stage. The primary was initially $6$--$14\,M_{\sun}$ and has lost its outer layers. This model predicts a $\sin i$ value close to 1, but no eclipses have been reported. A more detailed orbit was estimated by \citet{1988A&AS...75..157M}: $P=362.8\pm 0.1$ d, $K=48\pm 2$ \kms, $e=0.30\pm 0.03$. KS Per has been proposed as a binary BH in the literature \citep{1966ApJ...144..840Z, 1966SvA....10..251G}.

HIP 27655 = HD 270196 is a blue supergiant in the LMC, with little specific information in the literature. Its Gaia parallax may be underestimated but consistent with the distance to LMC within the formal error. 

HIP 37458 has also attracted little attention from observational astrophysicists. It is listed as variable Fp star in SIMBAD. It has a high proper motion in Gaia DR3, which, in combination with the small parallax, indicates a high space velocity of nearly 100 \kms. However, the RUWE parameter (5.3) is quite high, indicating a significant degree of perturbation in Gaia results. Gaia also determined a mean RV$=37.26$ \kms\ with a large amplitude of 20.3 \kms. \citet{2001BaltA..10..589A} reported a high degree of variability with an amplitude of 0.4 mag from the available Hipparcos light curve. We reanalyzed the Hipparcos light curve for this star using the unweighted least-squares periodogram and detected a clear periodic signal at 57 d, amplitude 0.17 mag, accompanied by smaller 1/2 and 1/4 harmonics. This type of photometric variability may be caused by strong tidal deformation of the primary from a massive orbiting companion.

HIP 78765 is an F7V star with an outstanding space velocity in excess of 100 \kms, if we believe the Gaia-determined values of parallax (0.27$\pm$0.49\,mas) and proper motion. Doubts emerge from a high RUWE value (22.1) and the inconsistency between the Gaia and Hipparcos astrometry. This is a known visual binary (WDS 16048$-$4044, I~1284) with an orbital period of 240\,yr \citep{2022AJ....164...58T}. The pair was not resolved by Hipparcos, being too close at the time.  Both the dynamical parallax derived from the orbit, about 6\,mas, and the Hipparcos parallax of 3.05$\pm$1.32 do not match the small and inaccurate Gaia parallax, explaining the large apparent velocity. In short,  another "star with complications" and obvious false positive. 

%The relative position of its known companion in the Washington Double Star catalog changed from $0.3\arcsec$ to $0.1\arcsec$ in separation and from $97\degr$ tp $60\degr$ in position angle between 1928 and 2018. This seems to be too fast for a distant star outside 1 kpc. \citet{2012AJ....144...56T} measured sep$=0.122\arcsec$ and PA$=234.9\degr$ (flipped companions?) for 2012.3539 and sep$=0.1387\arcsec$ and PA$=60.2\degr$ seven years later \citep{2019AJ....158...48T}. A gross error in Gaia parallax caused by unresolved duplicity may be suspected in this case.

HIP 88838 = VX Sgr is one of the most enigmatic objects in Table \ref{30.tab}. It has attracted much attention from observers due to its outstanding characteristics. It has been classified as a long-period variable star with one of the greatest amplitudes of variability. Indeed, the Gaia DR3 light curve varies between 8 and 10.5 mag in $G_{\rm BP}$. The star is flagged as variability-induced mover (VIM) in Hipparcos, which is generally associated with uncertain astrometry. The light curve pattern is closer to that of a Mira-type variable; however, there are periods of small photometric variability \citep{1972ApJ...172L..59H}. As expected for a Mira-variable, the light curves are different in different bands, and the characteristic period is approximately 732 d, but strong bands of CN identify the star as a Ia supergiant. The extreme IR radiation and a resolved maser source are explained in terms of an expanding circumstellar envelope. The shell is more than $1\arcsec$ in diameter \citep{1982MNRAS.198..385L}, so it should be huge given the estimated distance. \citet{2021A&A...646A..98T} argue that VX Sgr is more likely to be an extreme AGB star, or even a Thorne-\.Zytkow object with a degenerate star at the core. \citet{2010A&A...511A..51C} determined the angular diameter of the photosphere to be $8.82\pm 0.50$ mas. This makes the star improbably large if the Gaia DR3 parallax is correct. The Gaia DR3 astrometric values are likely to be off, because they are not consistent with the DR2 results (where $\varpi=0.787\pm 0.229$ mas), which are loosely consistent with the previously suggested membership of VX Sgr in the Sgr OB1 association in the Carina arm, as well as VLBI phase-referencing measurements by \citet{2018ApJ...859...14X} ($\varpi=0.64\pm0.04$). The large perturbation in Gaia DR3 astrometry is probably caused by the visible disk of this hypergiant star showing a clumpy and asymmetric structure \citep{2022A&A...658A.185C}, which is likely to vary on the scale of several mas due to the motion of bright spots.

\subsection{Correction of systematic errors}
\label{syst.sec}
The systematic errors of Hipparcos and Gaia astrometry are known to strongly correlate with the scanning schedules of these space missions. The genuine acceleration vectors, on the contrary, are expected to be randomly directed on the sky. This can be used to separate the systematic and physical components and subtract the sky-correlated error from the acceleration vectors. We use the vector spherical decomposition method \citep[VSH, ][]{2004ASPC..316..230V, 2007AJ....134..367M, 2009A&A...506.1477T, 2010AIPC.1283...94V, 2012A&A...547A..59M} on 75,686 Hipparcos-Gaia stars with $\chi_{\boldsymbol{\mu}}<5$, which effectively removes the binary stars with perturbed proper motion differences. The computation closely follows the algorithm described in \citep{2022AJ....164..157M}. 

The VSH fit is obtained by solving the linear problem
\eb 
\begin{split}
\boldsymbol{\mu}_{\rm G} & -\boldsymbol{\mu}_{\rm HG} = \sum_{l=1}^{L} [c_{0l0}\boldsymbol{S}_{0l0}(\alpha,\delta)+ d_{0l0}\boldsymbol{T}_{0l0}(\alpha,\delta) \\
&+\sum_{k=1}^2 \sum_{m=1}^l\; c_{klm}\boldsymbol{S}_{klm}(\alpha,\delta) +
d_{klm}\boldsymbol{T}_{klm}(\alpha,\delta) ],
\end{split}
\label{vsh.eq}
\ee
where $\boldsymbol{S}_{klm}(\alpha,\delta)$ and $\boldsymbol{T}_{klm}(\alpha,\delta)$
are, respectively, electric and magnetic VSH functions of celestial coordinates of real/imaginary kind $k$, degree $l$, and order $m$. This problem is solved by a weighted least-squares method with three-dimensional condition and weight matrices, since each condition equation defines a two-dimensional vector on the unit sphere. The resulting set of coefficients $c$ and $d$ define a vector field on the sphere, which represents the large-scale, or smoothly variable component of the observed proper motion field. The characteristic scale of the fitted variations is inversely related to the limiting degree $L$. The total number of fitting VSH is $2 L (L+2)$, and the lowest-degree fit with $L=1$ includes 6 VSH terms representing rigid rotation and dipole. For the chosen $L=7$, the number of fitting VSH terms is 126. The fit is a continuous vector-valued function, and a specific value can be computed for any point with coordinates $(\alpha,\delta)$.
The coefficients for the most significant terms are presented {\it ibid.}, Table 1, and the overall fit in the graphical form, in their Fig. 2. The fitted vector field is relatively small with a median length of 9 \uasyr, but the statistical significance of 52 VSH terms, quantified as the signal-to-noise ratio, is above 3.

The modal value of the formal uncertainty of $\Delta\boldsymbol{\mu}$ values is approximately 40 \uasyr, and the median of the fitted VSH field is 9 \uasyr. One-tenth of the general sample has the fitted vectors longer than 15 \uasyr. Thus, the systematic error is a non-negligible contributor to the observed $\Delta\boldsymbol{\mu}$ field. We subtract the fitted vectors for each of the general sample of 102,549 stars. Recomputing the $\delta_v$ values from the corrected HGPM vectors, we find only 14 stars with velocity difference above 30 km s$^{-1}$ against the previous 16 stars. This collection of surviving candidate BH binaries includes mostly stars in the southern hemisphere, with only three been in the north. All have  Gaia-determined parallaxes well below 1 mas, and it is possible that the $\delta_v$ signals are strongly overestimated because of the parallax uncertainty. Indeed, if we change the parallax values in Eq. \ref{deltav.eq} to $\varpi+3 \sigma_\varpi$, only one candidate still has $\delta_v>30$ \kms, which is the red supergiant VX Sgr. Although the probability of a $3\sigma$ error in parallax seems to be low for the general population of Gaia stars, we are investigating here a much smaller collection of peculiar  objects with unusual astrometric properties, which may represent the small fraction of extreme statistical outliers. Apart from the possibly incorrect Gaia parallaxes for these stars, they also have large angular sizes spanning milliarcseconds, and the intrinsic structural changes in their envelopes may produce spurious proper motions. However, other candidates are members of hierarchical multiple systems that have been astrometrically monitored, such as HIP 65385 \citep{2018ApJS..235....6T, 2020AJ....160....7T}.

\subsection{Bimodality of apparent accelerations and the search for astrometric exoplanets}
\begin{figure*}
\includegraphics[width=0.47 \textwidth]{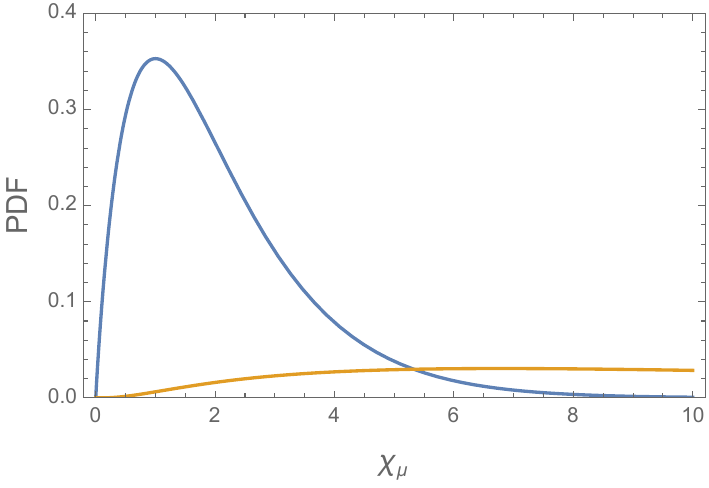}
\includegraphics[width=0.47 \textwidth]{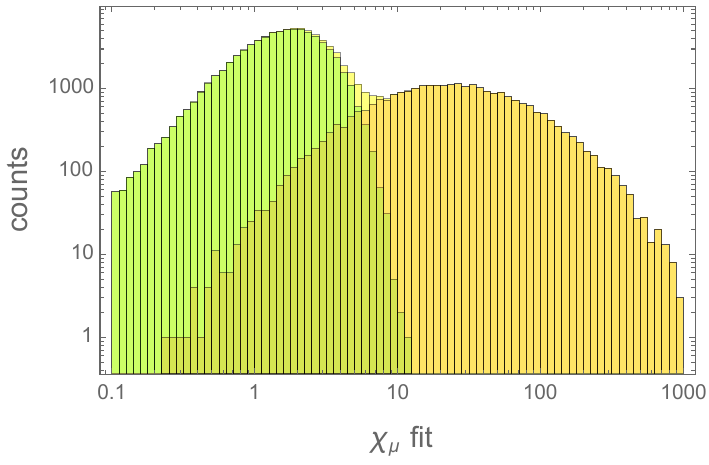}
\caption{Best-fitting mixed distribution of normalized apparent acceleration values $\chi_{\boldsymbol{\mu}}$ for 102,549 Hipparcos-Gaia stars. Left panel: PDF curves of the fitting mixed distribution. Right panel: corresponding histograms of the mixed distribution fit and its two components in the log-log scale simulated by random number generator, which can be compared with Fig. \ref{dmu.fig}, left.
\label{dist.fig}
}
\end{figure*}

The bimodal character of the acceleration signals $\delta_{\boldsymbol{\mu}}$ and their normalized magnitudes $\chi_{\boldsymbol{\mu}}$, illustrated in Fig. \ref{dmu.fig}, is also clearly seen when the VSH-corrected data are formally represented by arbitrary statistical distributions with free parameters. Using the Wolfram Mathematica function {\it FindDistribution}\footnote{\url{https://reference.wolfram.com/language/ref/FindDistribution.html}}, we find that the optimal fits for both sample distributions in this Figure  are sums of a Gamma distribution with two parameters, and a lognormal distribution. The proportions are: 72\% GammaDistribution[1.91, 0.044] and 28\% 
LogNormalDistribution[0.051, 1.20] for $\delta_{\boldsymbol{\mu}}$ (the scale is in mas), and 78\% GammaDistribution[1.93, 1.07] and 22\% LogNormalDistribution[3.08, 1.08] for $\chi_{\boldsymbol{\mu}}$. The PDF of the latter mixed distributions is shown in Fig. \ref{dist.fig}, left panel. To compare this fit with the observed distribution in Fig. \ref{dmu.fig}, left panel, we generated a sample of the same size using random number generation with the given distributions, and plotted the resulting histograms in logarithmic axes. The green shaded histogram corresponds to the GammaDistribution component, the orange-shaded part to the LogNormalDistribution component, and the combined histogram is shown in yellow.

It may be proposed that the second (log-normal) component of the distribution mostly includes astrometric binaries with massive (but faint or invisible) companions. Some of these stars have indeed been confirmed as such by spectroscopic or imaging observations. This population constitutes more than one-fifth of the general sample. The main component following the Gamma distribution is also abnormal from the statistical point of view. Although the distribution peaks at 1.07, which is close to the expected mode 1, the PDF is much different from the expected Rayleigh[1] distribution, being flatter at the top with a heavy tail stretching beyond $\chi_{\boldsymbol{\mu}}=3$, where the Rayleigh PDF becomes close to zero. This excess acceleration signal may be caused by planetary-mass companions in wide orbits around closer stars. We estimate from the analysis of the corresponding CDFs, that up to 40\% of this component, or roughly 35,000 stars may be in this category.

\section{Verification of accelerating binary HIP 44842} \label{section:HIP44842}

Statistical evidence of accelerated motion is not sufficient to identify {\it bona fide} binaries with massive and dim components, as we have seen on the example of the candidates with the largest $\delta_v$ signals. A dedicated spectroscopic and, possibly, imaging follow-up verification is required. Here we describe the principles of a proposed verification program on the example of a single previously selected candidate object. Since we have used a single spectroscopic instrument in a ``target of opportunity" mode, the selection was based on availability and practical convenience in this case. The investigated star is not the greatest signal ($\delta_v\approx 4.7$ \kms), but it is bright and accessible. Spectroscopic orbit determinations take time, and in this case, observations were initiated before the release of Gaia DR3. Our results help to verify Gaia data as well, since an orbital solution for the selected object was published in DR3.

The K2III giant HIP~44842 (HD~78788,  CD$-$50 3869, $V=8.55$ mag, $G=8.18$ mag) has a large acceleration signal in terms of the normalized value detected in this study. The corrected GPM$-$HGPM difference is $\Delta\boldsymbol{\mu}=(-0.427, -2.796)$ \masyr, with the formal $\chi_{\boldsymbol{\mu}}=14.53$, which is the 0.837 quantile of the sample. Thus, it certainly belongs to the lognormal component of the distribution, which certifies its astrometric binarity. It was detected as an accelerating star from the dedicated analysis of the Hipparcos mission data, which obtained an acceleration vector $(-15.5,-3.8)$ \masyrtwo\ with a high formal significance of 5.46. However, it was not selected in the review of candidate binaries with longer orbital periods based on the comparison of Tycho-2 and Hipparcos mean proper motions \citep{2005AJ....129.2420M}. This indicates a shorter orbital period of several years. The absence of a long-term signal was also confirmed by \citet{2007A&A...464..377F}.

However, the acceleration alone does not tell us much about the companion's mass without knowledge of the full orbit. The expected period of a few years makes this star a good target for spectroscopic follow-up. For this reason, we started to monitor the RV of this object in 2020. In 2022 June, the Gaia Non-Single Stars (NSS) catalog was released \citep{2023A&A...674A...9H}. It contained a spectro-astrometric orbit of HIP~44842 with a period of 1056.15 days and an amplitude of 3.63\,mas. This orbit matched our RV data, and here we combine all available information to evaluate the mass of the companion.

\subsection{CHIRON Spectroscopy}
\label{sec:chi}

High-resolution ($R=80,000$) spectra of HIP~44842 have been taken with the CHIRON optical echelle spectrometer \citep{CHIRON} installed at the 1.5-m telescope at Cerro Tololo in Chile. The telescope and instrument are operated by the   Small  and   Medium  Aperture  Telescopes   Research  System (SMARTS)
Consortium. %\footnote{ \url{http://www.astro.yale.edu/smarts/}}
The spectra were taken in the service mode and reduced by the pipeline.  The RVs are determined by cross-correlation of the reduced spectra with a binary mask based on the solar spectrum, as described by \citet{chiron1}. The cross-correlation functions (CCF) had a prominent narrow dip with an amplitude of 0.55 and an rms width of 4.13 \kms. Overall, we took eight spectra in the period from 2020.0 to 2022.9.

\begin{deluxetable*}{l ccc c ccc }    
\tabletypesize{\scriptsize}     
\tablecaption{Orbits of HIP 44842
\label{tab:sborb}          }
\tablewidth{0pt}                                 
\tablehead{                                                                     
\colhead{Source} & 
\colhead{$P$} & 
\colhead{$T$} & 
\colhead{$e$} & 
\colhead{$\omega_{\rm A}$ } & 
\colhead{$K_1$} & 
\colhead{$K_2$} & 
\colhead{$\gamma$} 
\\
& \colhead{(d)} &
\colhead{(JD -2,400,000)} & &
\colhead{(deg)} & 
\colhead{(km~s$^{-1}$)} &
\colhead{(km~s$^{-1}$)} &
\colhead{(km~s$^{-1}$)} 
}
\startdata
NSS     &  1056.2 & 57388.46  &   0.306  &     56.2 & 7.85 & \ldots & 62.55 \\
    & $\pm$6.4    & $\pm$3.5 &  $\pm$0.005    & \ldots & \ldots & \ldots &  $\pm$0.05 \\
CHIRON  & 1054.9   & 59216.7   &   0.314  &   236.0   & 8.65 & 9.48  &  63.31 \\
    & $\pm$8.5    & $\pm$5.1 &  $\pm$0.008 & $\pm$2.0 & $\pm$0.09 &$\pm$0.42 & $\pm$0.07 \\
\enddata 
\end{deluxetable*}

% radial velocities
\begin{deluxetable}{c c rrr  }    
\tabletypesize{\scriptsize}     
\tablecaption{Radial Velocities and Residuals
\label{tab:rv}          }
\tablewidth{0pt}                                   
\tablehead{                                                                     
\colhead{Date} & 
\colhead{Comp.} & 
\colhead{RV} & 
\colhead{$\sigma$} & 
\colhead{(O$-$C) } \\
\colhead{(JD -2,400,000)} & &
\multicolumn{3}{c}{(km s$^{-1}$)} 
}
\startdata
 59178.859 & a & 54.13  &   0.10&   $-$0.03  \\
 59178.859 & b & 73.25  &   0.70&   $-$0.21  \\
 59212.795 & a & 56.46  &   0.10&    0.03  \\
 59212.795 & b & 71.64  &   0.70&    0.70  \\
 59275.637 & a & 62.49  &   0.20&    0.05  \\
 59365.474 & a & 68.33  &   0.10&   $-$0.04  \\
 59365.474 & b & 57.82  &   0.70&   $-$0.02  \\
 59540.827 & a & 70.02  &   0.10&    0.08  \\
 59540.827 & b & 57.11  &   0.70&    1.00  \\
 59706.463 & a & 67.24  &   0.10&   $-$0.09  \\
 59706.463 & b & 58.09  &   0.70&   $-$0.89  \\
 59913.821 & a & 61.66  &   0.20&    0.07  \\
\enddata 
\end{deluxetable}

%---------------------------------------------------------
\subsection{Spectroscopic and Astrometric Orbits}
\label{sec:orb}

\begin{figure}
\includegraphics[width=0.47 \textwidth]{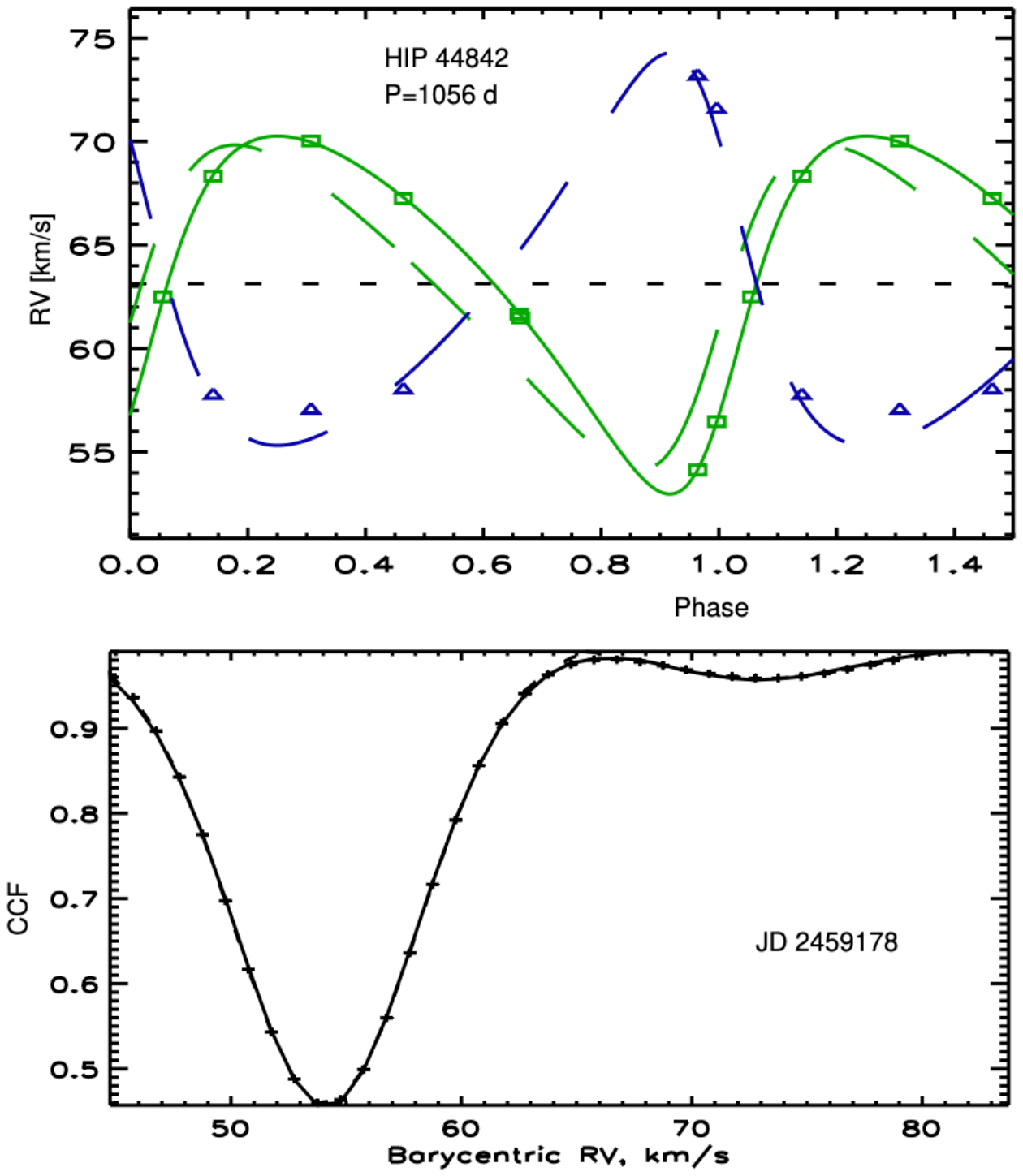}
\caption{Top panel: RV curve fitted to the CHIRON data (green solid line and
  squares for primary, blue dash and triangles for the secondary). The green dashed line shows the Gaia DR3 spectro-astrometric orbit. Bottom
  panel: the CCF recorded on JD 2459178.85 (line and pluses) fitted by
  a sum of two Gaussians (dash-dot line).
\label{fig:RV}
}
\end{figure}

The time span  of the Gaia DR3, 1020 days, is slightly  less than the
orbital period of 1056 days (2.8 yr).   The eight CHIRON  spectra cover an interval  of 1052
days. The  single-lined spectroscopic orbit fitted to our RVs is similar  to the
NSS spectro-astrometric  orbit. Assuming a  primary mass of  1.5 \msun
(see below), the orbits imply a secondary of $\sim$1.2 \msun.
The Gaia  spectro-astrometric orbit has  a node difference  of 180\degr
~compared to  CHIRON, as  all  such NSS  solutions.  The astrometric  Campbell
elements  derived  from  the  Thiele-Innes  parameters listed in the NSS  are:  $a  =
3.63$\,mas, $\Omega = 63\fdg5$, and $i=140\fdg2$.

If the  secondary companion of  HIP 44842 were a  normal main-sequence
star, it  would have an  absolute magnitude  $M_V \approx 4$  mag and
would contribute  about 6\% of  the flux in  the $V$ band,  making its
dips in  the CCF potentially  detectable.  Re-examination of  the CCFs
indeed revealed weak secondary dips shifted from the gamma-velocity
opposite to the main dip. We also noted  that the CCF dips  at phases where
the  lines  were  blended  were  a  few  percent  stronger,  confirming
indirectly that  the companion  is not  ``dark''. The  CCFs
were re-fitted assuming two dips, and  the resulting RVs were used to derive
the revised double-lined  orbit.  This orbit is plotted  in the top
panel  of Fig.~\ref{fig:RV},  while its  bottom panel  gives the  best
example of  a double dip approximated  by two Gaussians.  Elements  of the
NSS  and our  orbits  are listed  in  Table~\ref{tab:sborb}.  The  rms
residuals for  the primary and the  secondary are 0.06 and  0.68 \kms,
respectively. Individual RVs and residuals to the orbit are listed in Table~\ref{tab:rv}, where the main and secondary dips corresponds to the components a and b, respectively.  
In the least-square orbit fit the weights are inversely proportional to the RV errors, assumed to be 0.1 and 0.7 \kms for the main and secondary dips.
The areas of the CCF  dips are 2.19 and 0.15 \kms, their
ratio corresponds  to a magnitude  difference of 3.8  mag. 
Considering dependence of the  line contrast on effective temperature,
the actual magnitude difference could be a little  less.

%---------------------------------------------------------
\subsection{Parameters of the Stars}
\label{sec:par}

\begin{figure}
\includegraphics[width=0.37 \textwidth, angle=-90]{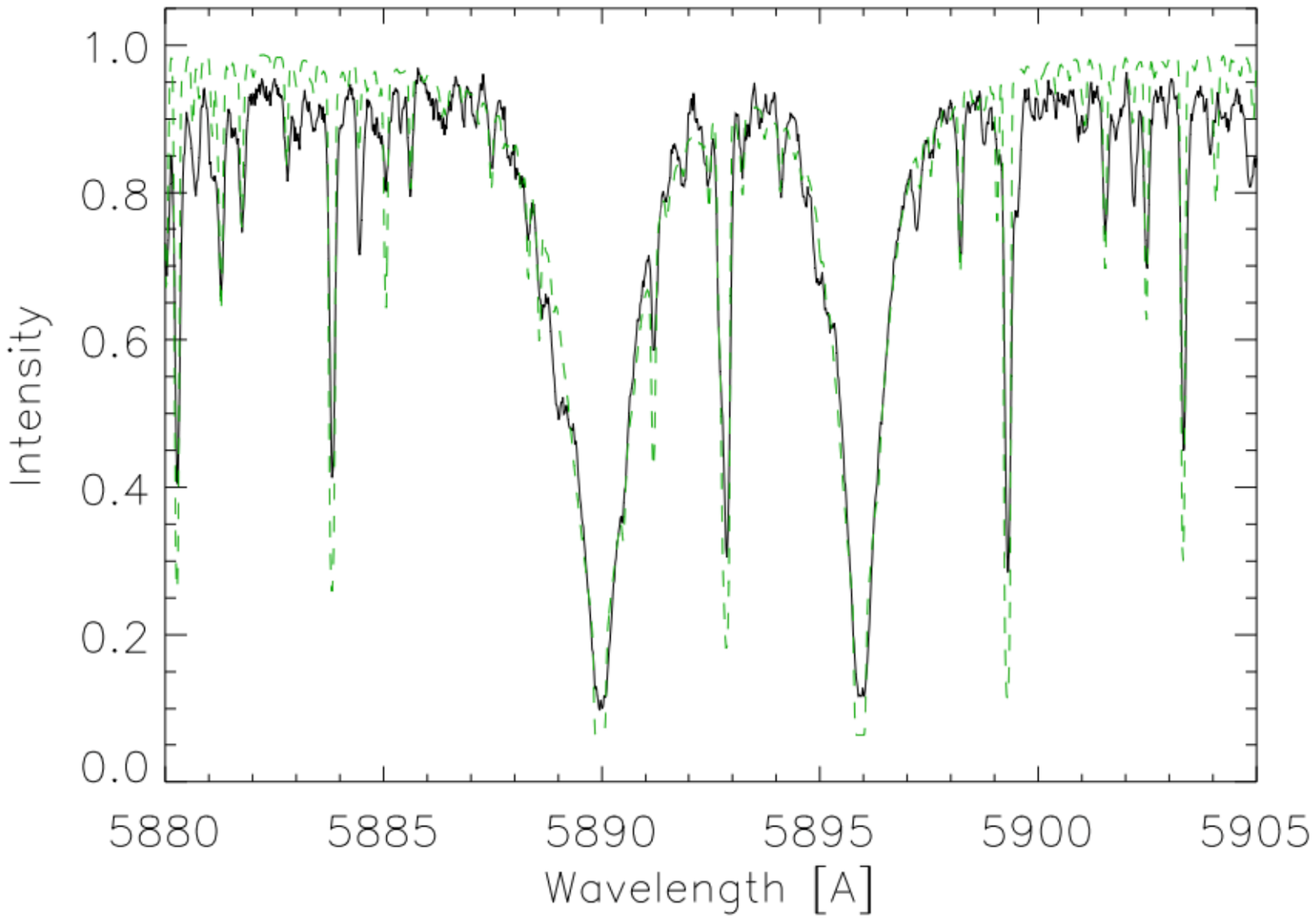}
\caption{Average spectrum of HIP 44842 shifted to zero velocity (full
  line) and the POLLUX synthetic spectrum \citep{2010A&A...516A..13P} for $T_e = 4500$\,K, $\log g = 3.0$, and  solar metallicity (green dash line). 
\label{fig:spec}
}
\end{figure}

\begin{figure}
\includegraphics[width=0.47 \textwidth]{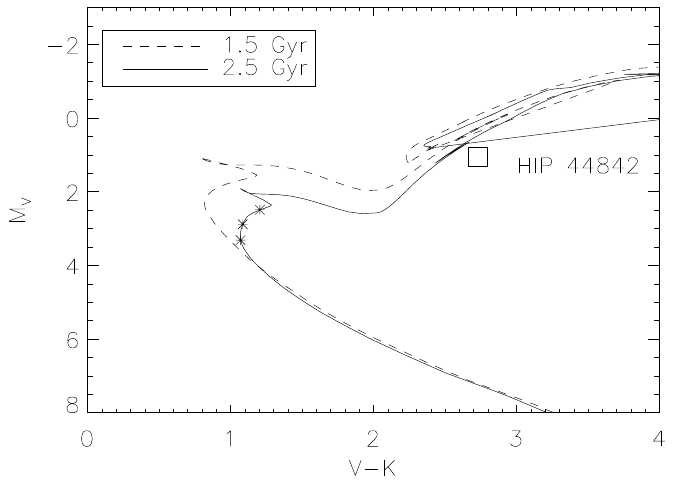}
\caption{PARSEC isochrones \citep{Bressan2012} for 1.5 and 2.5 Gyr ages and solar metallicity in the $(V-K, M_V)$  color-magnitude diagram. Asterisks mark masses of 1.2, 1.3, and 1.4
  \msun on the 2.5 Gyr isochrone.
\label{fig:CMD}
}
\end{figure}

The  Gaia parallax  fitted in  the NSS solution along with the astrometric orbit,
$3.171 \pm 0.018$ mas, corresponds to a distance modulus of 7.5 mag.
Figure~\ref{fig:CMD} plots  two isochrones from the  PARSEC collection \citep{Bressan2012} and compares  with  the position  of HIP~44842  at
$V-K=2.73$ mag and $M_V = 1.06$  mag (the $V-K$ color is consistent with $T_e
= 4500$\,K).  Although reliable estimates of the age cannot be derived
because the isochrones  are not monotonic in this region, it is likely  around
3\,Gyr.  The  mass of  the giant  is then about  1.5 \msun  or slightly
larger. The mass estimate is primarily constrained by the luminosity.

The double-lined orbit with a known inclination leads to the masses of 
1.11 and 1.0 \msun, less than inferred from the isochrones. A mass
sum of 2.1 \msun and a period of 2.89 yr correspond to a semimajor axis
of 2.61 au or 8.26 mas. The wobble factor $f$ (ratio of astrometric and
full axes) is 0.44. If the companion's light is neglected, the
estimated wobble factor $f = q/(1-q)=0.48$, where $q=0.91$ is the
spectroscopic mass ratio.  

Overall, there  is a broad  agreement between the  CHIRON double-lined
orbit and  the Gaia  NSS spectro-astrometric orbit.   The masses estimated
from the CHIRON  orbit are lower than inferred from  the isochrone by a
factor of  $\sim$1.4.  We note, however, that  the measured masses depend
on  the  orbital  inclination  $i$  as  $\sin^3  i$.  The  discrepancy
disappears  if   the  inclination   is  changed  from   140\fdg2  to
145\degr. Considering that Gaia DR3 covered only one orbital  cycle, the
NSS orbit might still be inaccurate.

%---------------------------------------------------------
\section{Summary}
\label{sec:summary}

Although the quest for discovering dormant StMBHs in relatively wide binaries has resulted in two first reliable detections \citep{2023MNRAS.518.1057E, 2023MNRAS.521.4323E}, it proved to be extremely difficult, with many failed efforts and false-positive cases. 
Here we performed an independent screening of Hipparcos stars for large accelerations by matching with Gaia DR3 and found several interesting cases. 

Astrometric acceleration emerges when the most accurate HGPM proper motions derived from the mean epoch positions of common stars in Hipparcos and Gaia are compared with the short-term Gaia proper motions. We improve these proper motion differences by removing systematic, sky-correlated errors caused mostly by Hipparcos position systematics, using the VSH fitting up to degree 7. Although the median correction is just 9 \uasyr, it proves to clean out some of the false positives with the smallest $\Delta\boldsymbol{\mu}$ signals, including candidate dark companions. Selecting a working sample of stars with the velocity differences above 30 \kms, we find 14 candidates with small Gaia-determined parallaxes. Provisional analysis of some of these candidates shows that they may all be false positives caused by a range of complications, most commonly, an incorrect or imprecise Gaia parallax. Indeed, if the parallaxes are increased by $3\,\sigma_\varpi$, all these examples fall below the threshold 30 \kms\ with the exception of VX Sgr.

Our study is further focused on less exotic objects with more reliable parallaxes but still significant acceleration estimates. We started spectroscopic monitoring of the large-acceleration star HIP~44842 before the release of Gaia DR3 in search of a potentially massive and dark  companion. Our spectroscopy eventually revealed a  small signature of this companion in the spectra, proving that it is not a compact object. Another common type of massive faint companions,  namely  close  pairs of low-mass  stars,  is  also rejected in  this case because we detected the lines of the secondary.

A cleaner sample of candidate accelerating stars with massive invisible companions will be achieved after the next data release of Gaia (DR4).
The most important aspect of improvement concerns the calibration techniques  for the brighter stars from the Hipparcos sample, which are problematic for Gaia \citep{2021A&A...649A...2L}. If the Hipparcos-Gaia collection of 100,000 stars proves to be too small to find StMBH binaries, the longer time span of accurate epoch astrometry should facilitate detection of orbiting pairs with periods up to several years \citep{2023A&A...674A..34G}.

\section*{Acknowledgements}
We thank the operators of the 1.5-m telescope for executing observations of
this  program  and  the   SMARTS  team  for  scheduling  and  pipeline
processing.
The research was funded by the NSF's NOIRLab and by USNO.
This work  used the  SIMBAD service operated  by Centre  des Donn\'ees
Stellaires  (Strasbourg, France) and  bibliographic  references from  the
Astrophysics Data  System maintained  by SAO/NASA.
This work has made use of data from the European Space Agency (ESA) mission
{\it Gaia} (\url{https://www.cosmos.esa.int/gaia}), processed by the {\it Gaia}
Data Processing and Analysis Consortium (DPAC,
\url{https://www.cosmos.esa.int/web/gaia/dpac/consortium}). Funding for the DPAC
has been provided by national institutions, in particular the institutions
participating in the {\it Gaia} Multilateral Agreement.

\facilities{CTIO:1.5m, Gaia}

%\software{Astropy \citep{2013A&A...558A..33A, 2018AJ....156..123A}, \textsc{topcat} \citep{2005ASPC..347...29T}}

\bibliography{enigma}

\begin{thebibliography}{}
\expandafter\ifx\csname natexlab\endcsname\relax\def\natexlab#1{#1}\fi
\providecommand{\url}[1]{\href{#1}{#1}}
\providecommand{\dodoi}[1]{doi:~\href{http://doi.org/#1}{\nolinkurl{#1}}}
\providecommand{\doeprint}[1]{\href{http://ascl.net/#1}{\nolinkurl{http://ascl.net/#1}}}
\providecommand{\doarXiv}[1]{\href{https://arxiv.org/abs/#1}{\nolinkurl{https://arxiv.org/abs/#1}}}

\bibitem[{{Adelman}(2001)}]{2001BaltA..10..589A}
{Adelman}, S.~J. 2001, Baltic Astronomy, 10, 589,
  \dodoi{10.1515/astro-2001-0403}

\bibitem[{{Bidelman}(1950)}]{1950ApJ...111..333B}
{Bidelman}, W.~P. 1950, \apj, 111, 333, \dodoi{10.1086/145268}

\bibitem[{{Brandt}(2021)}]{Brandt2021}
{Brandt}, T.~D. 2021, \apjs, 254, 42, \dodoi{10.3847/1538-4365/abf93c}

\bibitem[{{Bressan} {et~al.}(2012){Bressan}, {Marigo}, {Girardi}, {Salasnich},
  {Dal Cero}, {Rubele}, \& {Nanni}}]{Bressan2012}
{Bressan}, A., {Marigo}, P., {Girardi}, L., {et~al.} 2012, \mnras, 427, 127,
  \dodoi{10.1111/j.1365-2966.2012.21948.x}

\bibitem[{{Butkevich} \& {Lindegren}(2014)}]{2014A&A...570A..62B}
{Butkevich}, A.~G., \& {Lindegren}, L. 2014, \aap, 570, A62,
  \dodoi{10.1051/0004-6361/201424483}

\bibitem[{{Casares} {et~al.}(1992){Casares}, {Charles}, \&
  {Naylor}}]{1992Natur.355..614C}
{Casares}, J., {Charles}, P.~A., \& {Naylor}, T. 1992, \nat, 355, 614,
  \dodoi{10.1038/355614a0}

\bibitem[{{Chiavassa} {et~al.}(2010){Chiavassa}, {Lacour}, {Millour}, {Driebe},
  {Wittkowski}, {Plez}, {Thi{\'e}baut}, {Josselin}, {Freytag}, {Scholz}, \&
  {Haubois}}]{2010A&A...511A..51C}
{Chiavassa}, A., {Lacour}, S., {Millour}, F., {et~al.} 2010, \aap, 511, A51,
  \dodoi{10.1051/0004-6361/200913288}

\bibitem[{{Chiavassa} {et~al.}(2022){Chiavassa}, {Kravchenko}, {Montarg{\`e}s},
  {Millour}, {Matter}, {Freytag}, {Wittkowski}, {Hocd{\'e}}, {Cruzal{\`e}bes},
  {Allouche}, {Lopez}, {Lagarde}, {Petrov}, {Meilland}, {Robbe-Dubois},
  {Hofmann}, {Weigelt}, {Berio}, {Bendjoya}, {Bettonvil}, {Domiciano de Souza},
  {Heininger}, {Henning}, {Isbell}, {Jaffe}, {Labadie}, {Lehmitz},
  {Meisenheimer}, {Soulain}, {Varga}, {Augereau}, {van Boekel}, {Burtscher},
  {Danchi}, {Dominik}, {Drevon}, {G{\'a}mez Rosas}, {Hogerheijde}, {Hron},
  {Klarmann}, {Kokoulina}, {Lagadec}, {Leftley}, {Mosoni}, {Nardetto},
  {Paladini}, {Pantin}, {Schertl}, {Stee}, {Szabados}, {Waters}, {Wolf}, \&
  {Yoffe}}]{2022A&A...658A.185C}
{Chiavassa}, A., {Kravchenko}, K., {Montarg{\`e}s}, M., {et~al.} 2022, \aap,
  658, A185, \dodoi{10.1051/0004-6361/202142514}

\bibitem[{{Drilling} \& {Schonberner}(1982)}]{1982A&A...113L..22D}
{Drilling}, J.~S., \& {Schonberner}, D. 1982, \aap, 113, L22

\bibitem[{{El-Badry} {et~al.}(2023{\natexlab{a}}){El-Badry}, {Rix}, {Quataert},
  {Howard}, {Isaacson}, {Fuller}, {Hawkins}, {Breivik}, {Wong}, {Rodriguez},
  {Conroy}, {Shahaf}, {Mazeh}, {Arenou}, {Burdge}, {Bashi}, {Faigler}, {Weisz},
  {Seeburger}, {Almada Monter}, \& {Wojno}}]{2023MNRAS.518.1057E}
{El-Badry}, K., {Rix}, H.-W., {Quataert}, E., {et~al.} 2023{\natexlab{a}},
  \mnras, 518, 1057, \dodoi{10.1093/mnras/stac3140}

\bibitem[{{El-Badry} {et~al.}(2023{\natexlab{b}}){El-Badry}, {Rix}, {Cendes},
  {Rodriguez}, {Conroy}, {Quataert}, {Hawkins}, {Zari}, {Hobson}, {Breivik},
  {Rau}, {Berger}, {Shahaf}, {Seeburger}, {Burdge}, {Latham}, {Buchhave},
  {Bieryla}, {Bashi}, {Mazeh}, \& {Faigler}}]{2023MNRAS.521.4323E}
{El-Badry}, K., {Rix}, H.-W., {Cendes}, Y., {et~al.} 2023{\natexlab{b}},
  \mnras, 521, 4323, \dodoi{10.1093/mnras/stad799}

\bibitem[{{Frankowski} {et~al.}(2007){Frankowski}, {Jancart}, \&
  {Jorissen}}]{2007A&A...464..377F}
{Frankowski}, A., {Jancart}, S., \& {Jorissen}, A. 2007, \aap, 464, 377,
  \dodoi{10.1051/0004-6361:20065526}

\bibitem[{{Gaia Collaboration} {et~al.}(2021){Gaia Collaboration}, {Brown},
  {Vallenari}, {Prusti}, {de Bruijne}, {Babusiaux}, {Biermann}, {Creevey},
  {Evans}, {Eyer}, {Hutton}, {Jansen}, {Jordi}, {Klioner}, {Lammers},
  {Lindegren}, {Luri}, {Mignard}, {Panem}, {Pourbaix}, {Randich}, {Sartoretti},
  {Soubiran}, {Walton}, {Arenou}, {Bailer-Jones}, {Bastian}, {Cropper},
  {Drimmel}, {Katz}, {Lattanzi}, {van Leeuwen}, {Bakker}, {Cacciari},
  {Casta{\~n}eda}, {De Angeli}, {Ducourant}, {Fabricius}, {Fouesneau},
  {Fr{\'e}mat}, {Guerra}, {Guerrier}, {Guiraud}, {Jean-Antoine Piccolo},
  {Masana}, {Messineo}, {Mowlavi}, {Nicolas}, {Nienartowicz}, {Pailler},
  {Panuzzo}, {Riclet}, {Roux}, {Seabroke}, {Sordo}, {Tanga}, {Th{\'e}venin},
  {Gracia-Abril}, {Portell}, {Teyssier}, {Altmann}, {Andrae}, {Bellas-Velidis},
  {Benson}, {Berthier}, {Blomme}, {Brugaletta}, {Burgess}, {Busso}, {Carry},
  {Cellino}, {Cheek}, {Clementini}, {Damerdji}, {Davidson}, {Delchambre},
  {Dell'Oro}, {Fern{\'a}ndez-Hern{\'a}ndez}, {Galluccio}, {Garc{\'\i}a-Lario},
  {Garcia-Reinaldos}, {Gonz{\'a}lez-N{\'u}{\~n}ez}, {Gosset}, {Haigron},
  {Halbwachs}, {Hambly}, {Harrison}, {Hatzidimitriou}, {Heiter},
  {Hern{\'a}ndez}, {Hestroffer}, {Hodgkin}, {Holl}, {Jan{\ss}en}, {Jevardat de
  Fombelle}, {Jordan}, {Krone-Martins}, {Lanzafame}, {L{\"o}ffler}, {Lorca},
  {Manteiga}, {Marchal}, {Marrese}, {Moitinho}, {Mora}, {Muinonen}, {Osborne},
  {Pancino}, {Pauwels}, {Petit}, {Recio-Blanco}, {Richards}, {Riello},
  {Rimoldini}, {Robin}, {Roegiers}, {Rybizki}, {Sarro}, {Siopis}, {Smith},
  {Sozzetti}, {Ulla}, {Utrilla}, {van Leeuwen}, {van Reeven}, {Abbas}, {Abreu
  Aramburu}, {Accart}, {Aerts}, {Aguado}, {Ajaj}, {Altavilla}, {{\'A}lvarez},
  {{\'A}lvarez Cid-Fuentes}, {Alves}, {Anderson}, {Anglada Varela}, {Antoja},
  {Audard}, {Baines}, {Baker}, {Balaguer-N{\'u}{\~n}ez}, {Balbinot}, {Balog},
  {Barache}, {Barbato}, {Barros}, {Barstow}, {Bartolom{\'e}}, {Bassilana},
  {Bauchet}, {Baudesson-Stella}, {Becciani}, {Bellazzini}, {Bernet}, {Bertone},
  {Bianchi}, {Blanco-Cuaresma}, {Boch}, {Bombrun}, {Bossini}, {Bouquillon},
  {Bragaglia}, {Bramante}, {Breedt}, {Bressan}, {Brouillet}, {Bucciarelli},
  {Burlacu}, {Busonero}, {Butkevich}, {Buzzi}, {Caffau}, {Cancelliere},
  {C{\'a}novas}, {Cantat-Gaudin}, {Carballo}, {Carlucci}, {Carnerero},
  {Carrasco}, {Casamiquela}, {Castellani}, {Castro-Ginard}, {Castro Sampol},
  {Chaoul}, {Charlot}, {Chemin}, {Chiavassa}, {Cioni}, {Comoretto}, {Cooper},
  {Cornez}, {Cowell}, {Crifo}, {Crosta}, {Crowley}, {Dafonte}, {Dapergolas},
  {David}, {David}, {de Laverny}, {De Luise}, {De March}, {De Ridder}, {de
  Souza}, {de Teodoro}, {de Torres}, {del Peloso}, {del Pozo}, {Delbo},
  {Delgado}, {Delgado}, {Delisle}, {Di Matteo}, {Diakite}, {Diener},
  {Distefano}, {Dolding}, {Eappachen}, {Edvardsson}, {Enke}, {Esquej}, {Fabre},
  {Fabrizio}, {Faigler}, {Fedorets}, {Fernique}, {Fienga}, {Figueras},
  {Fouron}, {Fragkoudi}, {Fraile}, {Franke}, {Gai}, {Garabato},
  {Garcia-Gutierrez}, {Garc{\'\i}a-Torres}, {Garofalo}, {Gavras}, {Gerlach},
  {Geyer}, {Giacobbe}, {Gilmore}, {Girona}, {Giuffrida}, {Gomel}, {Gomez},
  {Gonzalez-Santamaria}, {Gonz{\'a}lez-Vidal}, {Granvik},
  {Guti{\'e}rrez-S{\'a}nchez}, {Guy}, {Hauser}, {Haywood}, {Helmi}, {Hidalgo},
  {Hilger}, {H{\l}adczuk}, {Hobbs}, {Holland}, {Huckle}, {Jasniewicz},
  {Jonker}, {Juaristi Campillo}, {Julbe}, {Karbevska}, {Kervella}, {Khanna},
  {Kochoska}, {Kontizas}, {Kordopatis}, {Korn}, {Kostrzewa-Rutkowska},
  {Kruszy{\'n}ska}, {Lambert}, {Lanza}, {Lasne}, {Le Campion}, {Le Fustec},
  {Lebreton}, {Lebzelter}, {Leccia}, {Leclerc}, {Lecoeur-Taibi}, {Liao},
  {Licata}, {Lindstr{\o}m}, {Lister}, {Livanou}, {Lobel}, {Madrero Pardo},
  {Managau}, {Mann}, {Marchant}, {Marconi}, {Marcos Santos}, {Marinoni},
  {Marocco}, {Marshall}, {Martin Polo}, {Mart{\'\i}n-Fleitas}, {Masip},
  {Massari}, {Mastrobuono-Battisti}, {Mazeh}, {McMillan}, {Messina},
  {Michalik}, {Millar}, {Mints}, {Molina}, {Molinaro}, {Moln{\'a}r},
  {Montegriffo}, {Mor}, {Morbidelli}, {Morel}, {Morris}, {Mulone}, {Munoz},
  {Muraveva}, {Murphy}, {Musella}, {Noval}, {Ord{\'e}novic}, {Orr{\`u}},
  {Osinde}, {Pagani}, {Pagano}, {Palaversa}, {Palicio}, {Panahi}, {Pawlak},
  {Pe{\~n}alosa Esteller}, {Penttil{\"a}}, {Piersimoni}, {Pineau}, {Plachy},
  {Plum}, {Poggio}, {Poretti}, {Poujoulet}, {Pr{\v{s}}a}, {Pulone}, {Racero},
  {Ragaini}, {Rainer}, {Raiteri}, {Rambaux}, {Ramos}, {Ramos-Lerate}, {Re
  Fiorentin}, {Regibo}, {Reyl{\'e}}, {Ripepi}, {Riva}, {Rixon}, {Robichon},
  {Robin}, {Roelens}, {Rohrbasser}, {Romero-G{\'o}mez}, {Rowell}, {Royer},
  {Rybicki}, {Sadowski}, {Sagrist{\`a} Sell{\'e}s}, {Sahlmann}, {Salgado},
  {Salguero}, {Samaras}, {Sanchez Gimenez}, {Sanna}, {Santove{\~n}a},
  {Sarasso}, {Schultheis}, {Sciacca}, {Segol}, {Segovia}, {S{\'e}gransan},
  {Semeux}, {Shahaf}, {Siddiqui}, {Siebert}, {Siltala}, {Slezak}, {Smart},
  {Solano}, {Solitro}, {Souami}, {Souchay}, {Spagna}, {Spoto}, {Steele},
  {Steidelm{\"u}ller}, {Stephenson}, {S{\"u}veges}, {Szabados}, {Szegedi-Elek},
  {Taris}, {Tauran}, {Taylor}, {Teixeira}, {Thuillot}, {Tonello}, {Torra},
  {Torra}, {Turon}, {Unger}, {Vaillant}, {van Dillen}, {Vanel}, {Vecchiato},
  {Viala}, {Vicente}, {Voutsinas}, {Weiler}, {Wevers}, {Wyrzykowski}, {Yoldas},
  {Yvard}, {Zhao}, {Zorec}, {Zucker}, {Zurbach}, \& {Zwitter}}]{Gaia3}
{Gaia Collaboration}, {Brown}, A.~G.~A., {Vallenari}, A., {et~al.} 2021, \aap,
  649, A1, \dodoi{10.1051/0004-6361/202039657}

\bibitem[{{Gaia Collaboration} {et~al.}(2023){Gaia Collaboration}, {Arenou},
  {Babusiaux}, {Barstow}, {Faigler}, {Jorissen}, {Kervella}, {Mazeh},
  {Mowlavi}, {Panuzzo}, {Sahlmann}, {Shahaf}, {Sozzetti}, {Bauchet},
  {Damerdji}, {Gavras}, {Giacobbe}, {Gosset}, {Halbwachs}, {Holl}, {Lattanzi},
  {Leclerc}, {Morel}, {Pourbaix}, {Re Fiorentin}, {Sadowski}, {S{\'e}gransan},
  {Siopis}, {Teyssier}, {Zwitter}, {Planquart}, {Brown}, {Vallenari}, {Prusti},
  {de Bruijne}, {Biermann}, {Creevey}, {Ducourant}, {Evans}, {Eyer}, {Guerra},
  {Hutton}, {Jordi}, {Klioner}, {Lammers}, {Lindegren}, {Luri}, {Mignard},
  {Panem}, {Randich}, {Sartoretti}, {Soubiran}, {Tanga}, {Walton},
  {Bailer-Jones}, {Bastian}, {Drimmel}, {Jansen}, {Katz}, {van Leeuwen},
  {Bakker}, {Cacciari}, {Casta{\~n}eda}, {De Angeli}, {Fabricius}, {Fouesneau},
  {Fr{\'e}mat}, {Galluccio}, {Guerrier}, {Heiter}, {Masana}, {Messineo},
  {Nicolas}, {Nienartowicz}, {Pailler}, {Riclet}, {Roux}, {Seabroke}, {Sordo},
  {Th{\'e}venin}, {Gracia-Abril}, {Portell}, {Altmann}, {Andrae}, {Audard},
  {Bellas-Velidis}, {Benson}, {Berthier}, {Blomme}, {Burgess}, {Busonero},
  {Busso}, {C{\'a}novas}, {Carry}, {Cellino}, {Cheek}, {Clementini},
  {Davidson}, {de Teodoro}, {Nu{\~n}ez Campos}, {Delchambre}, {Dell'Oro},
  {Esquej}, {Fern{\'a}ndez-Hern{\'a}ndez}, {Fraile}, {Garabato},
  {Garc{\'\i}a-Lario}, {Haigron}, {Hambly}, {Harrison}, {Hern{\'a}ndez},
  {Hestroffer}, {Hodgkin}, {Jan{\ss}en}, {Jevardat de Fombelle}, {Jordan},
  {Krone-Martins}, {Lanzafame}, {L{\"o}ffler}, {Marchal}, {Marrese},
  {Moitinho}, {Muinonen}, {Osborne}, {Pancino}, {Pauwels}, {Recio-Blanco},
  {Reyl{\'e}}, {Riello}, {Rimoldini}, {Roegiers}, {Rybizki}, {Sarro}, {Smith},
  {Utrilla}, {van Leeuwen}, {Abbas}, {{\'A}brah{\'a}m}, {Abreu Aramburu},
  {Aerts}, {Aguado}, {Ajaj}, {Aldea-Montero}, {Altavilla}, {{\'A}lvarez},
  {Alves}, {Anders}, {Anderson}, {Anglada Varela}, {Antoja}, {Baines}, {Baker},
  {Balaguer-N{\'u}{\~n}ez}, {Balbinot}, {Balog}, {Barache}, {Barbato},
  {Barros}, {Bartolom{\'e}}, {Bassilana}, {Becciani}, {Bellazzini},
  {Berihuete}, {Bernet}, {Bertone}, {Bianchi}, {Binnenfeld}, {Blanco-Cuaresma},
  {Blazere}, {Boch}, {Bombrun}, {Bossini}, {Bouquillon}, {Bragaglia},
  {Bramante}, {Breedt}, {Bressan}, {Brouillet}, {Brugaletta}, {Bucciarelli},
  {Burlacu}, {Butkevich}, {Buzzi}, {Caffau}, {Cancelliere}, {Cantat-Gaudin},
  {Carballo}, {Carlucci}, {Carnerero}, {Carrasco}, {Casamiquela}, {Castellani},
  {Castro-Ginard}, {Chaoul}, {Charlot}, {Chemin}, {Chiaramida}, {Chiavassa},
  {Chornay}, {Comoretto}, {Contursi}, {Cooper}, {Cornez}, {Cowell}, {Crifo},
  {Cropper}, {Crosta}, {Crowley}, {Dafonte}, {Dapergolas}, {David}, {de
  Laverny}, {De Luise}, {De March}, {De Ridder}, {de Souza}, {de Torres}, {del
  Peloso}, {del Pozo}, {Delbo}, {Delgado}, {Delisle}, {Demouchy},
  {Dharmawardena}, {Diakite}, {Diener}, {Distefano}, {Dolding}, {Enke},
  {Fabre}, {Fabrizio}, {Fedorets}, {Fernique}, {Figueras}, {Fournier},
  {Fouron}, {Fragkoudi}, {Gai}, {Garcia-Gutierrez}, {Garcia-Reinaldos},
  {Garc{\'\i}a-Torres}, {Garofalo}, {Gavel}, {Gerlach}, {Geyer}, {Gilmore},
  {Girona}, {Giuffrida}, {Gomel}, {Gomez}, {Gonz{\'a}lez-N{\'u}{\~n}ez},
  {Gonz{\'a}lez-Santamar{\'\i}a}, {Gonz{\'a}lez-Vidal}, {Granvik}, {Guillout},
  {Guiraud}, {Guti{\'e}rrez-S{\'a}nchez}, {Guy}, {Hatzidimitriou}, {Hauser},
  {Haywood}, {Helmer}, {Helmi}, {Sarmiento}, {Hidalgo}, {Hilger},
  {H{\l}adczuk}, {Hobbs}, {Holland}, {Huckle}, {Jardine}, {Jasniewicz},
  {Jean-Antoine Piccolo}, {Jim{\'e}nez-Arranz}, {Juaristi Campillo}, {Julbe},
  {Karbevska}, {Khanna}, {Kordopatis}, {Korn}, {K{\'o}sp{\'a}l},
  {Kostrzewa-Rutkowska}, {Kruszy{\'n}ska}, {Kun}, {Laizeau}, {Lambert},
  {Lanza}, {Lasne}, {Le Campion}, {Lebreton}, {Lebzelter}, {Leccia},
  {Lecoeur-Taibi}, {Liao}, {Licata}, {Lindstr{\o}m}, {Lister}, {Livanou},
  {Lobel}, {Lorca}, {Loup}, {Madrero Pardo}, {Magdaleno Romeo}, {Managau},
  {Mann}, {Manteiga}, {Marchant}, {Marconi}, {Marcos}, {Marcos Santos},
  {Mar{\'\i}n Pina}, {Marinoni}, {Marocco}, {Marshall}, {Martin Polo},
  {Mart{\'\i}n-Fleitas}, {Marton}, {Mary}, {Masip}, {Massari},
  {Mastrobuono-Battisti}, {McMillan}, {Messina}, {Michalik}, {Millar}, {Mints},
  {Molina}, {Molinaro}, {Moln{\'a}r}, {Monari}, {Mongui{\'o}}, {Montegriffo},
  {Montero}, {Mor}, {Mora}, {Morbidelli}, {Morris}, {Muraveva}, {Murphy},
  {Musella}, {Nagy}, {Noval}, {Oca{\~n}a}, {Ogden}, {Ordenovic}, {Osinde},
  {Pagani}, {Pagano}, {Palaversa}, {Palicio}, {Pallas-Quintela}, {Panahi},
  {Payne-Wardenaar}, {Pe{\~n}alosa Esteller}, {Penttil{\"a}}, {Pichon},
  {Piersimoni}, {Pineau}, {Plachy}, {Plum}, {Poggio}, {Pr{\v{s}}a}, {Pulone},
  {Racero}, {Ragaini}, {Rainer}, {Raiteri}, {Ramos}, {Ramos-Lerate}, {Regibo},
  {Richards}, {Rios Diaz}, {Ripepi}, {Riva}, {Rix}, {Rixon}, {Robichon},
  {Robin}, {Robin}, {Roelens}, {Rogues}, {Rohrbasser}, {Romero-G{\'o}mez},
  {Rowell}, {Royer}, {Ruz Mieres}, {Rybicki}, {S{\'a}ez N{\'u}{\~n}ez},
  {Sagrist{\`a} Sell{\'e}s}, {Salguero}, {Samaras}, {Sanchez Gimenez}, {Sanna},
  {Santove{\~n}a}, {Sarasso}, {Schultheis}, {Sciacca}, {Segol}, {Segovia},
  {Semeux}, {Siddiqui}, {Siebert}, {Siltala}, {Silvelo}, {Slezak}, {Slezak},
  {Smart}, {Snaith}, {Solano}, {Solitro}, {Souami}, {Souchay}, {Spagna},
  {Spina}, {Spoto}, {Steele}, {Steidelm{\"u}ller}, {Stephenson}, {S{\"u}veges},
  {Surdej}, {Szabados}, {Szegedi-Elek}, {Taris}, {Taylor}, {Teixeira},
  {Tolomei}, {Tonello}, {Torra}, {Torra}, {Torralba Elipe}, {Trabucchi},
  {Tsounis}, {Turon}, {Ulla}, {Unger}, {Vaillant}, {van Dillen}, {van Reeven},
  {Vanel}, {Vecchiato}, {Viala}, {Vicente}, {Voutsinas}, {Weiler}, {Wevers},
  {Wyrzykowski}, {Yoldas}, {Yvard}, {Zhao}, {Zorec}, \&
  {Zucker}}]{2023A&A...674A..34G}
{Gaia Collaboration}, {Arenou}, F., {Babusiaux}, C., {et~al.} 2023, \aap, 674,
  A34, \dodoi{10.1051/0004-6361/202243782}

\bibitem[{{Goldin} \& {Makarov}(2006)}]{2006ApJS..166..341G}
{Goldin}, A., \& {Makarov}, V.~V. 2006, \apjs, 166, 341, \dodoi{10.1086/505939}

\bibitem[{{Goldin} \& {Makarov}(2007)}]{2007ApJS..173..137G}
---. 2007, \apjs, 173, 137, \dodoi{10.1086/520513}

\bibitem[{{Gould} \& {Salim}(2002)}]{2002ApJ...572..944G}
{Gould}, A., \& {Salim}, S. 2002, \apj, 572, 944, \dodoi{10.1086/340435}

\bibitem[{{Guseinov} \& {Zel'dovich}(1966)}]{1966SvA....10..251G}
{Guseinov}, O.~K., \& {Zel'dovich}, Y.~B. 1966, \sovast, 10, 251

\bibitem[{{Halbwachs} {et~al.}(2023){Halbwachs}, {Pourbaix}, {Arenou},
  {Galluccio}, {Guillout}, {Bauchet}, {Marchal}, {Sadowski}, \&
  {Teyssier}}]{2023A&A...674A...9H}
{Halbwachs}, J.-L., {Pourbaix}, D., {Arenou}, F., {et~al.} 2023, \aap, 674, A9,
  \dodoi{10.1051/0004-6361/202243969}

\bibitem[{{Heard} \& {Boshko}(1955)}]{1955AJ.....60..162H}
{Heard}, J.~F., \& {Boshko}, O. 1955, \aj, 60, 162, \dodoi{10.1086/107194}

\bibitem[{{H{\o}g} {et~al.}(2000){H{\o}g}, {Fabricius}, {Makarov}, {Bastian},
  {Schwekendiek}, {Wicenec}, {Urban}, {Corbin}, \&
  {Wycoff}}]{2000A&A...357..367H}
{H{\o}g}, E., {Fabricius}, C., {Makarov}, V.~V., {et~al.} 2000, \aap, 357, 367

\bibitem[{{Humphreys} \& {Lockwood}(1972)}]{1972ApJ...172L..59H}
{Humphreys}, R.~M., \& {Lockwood}, G.~W. 1972, \apjl, 172, L59,
  \dodoi{10.1086/180891}

\bibitem[{{Janssens} {et~al.}(2023{\natexlab{a}}){Janssens}, {Shenar},
  {Degenaar}, {Bodensteiner}, {Sana}, {Audenaert}, \& {Frost}}]{Janssens2023b}
{Janssens}, S., {Shenar}, T., {Degenaar}, N., {et~al.} 2023{\natexlab{a}},
  \aap, 677, L9, \dodoi{10.1051/0004-6361/202347318}

\bibitem[{{Janssens} {et~al.}(2023{\natexlab{b}}){Janssens}, {Shenar}, {Sana},
  \& {Marchant}}]{Janssens2023}
{Janssens}, S., {Shenar}, T., {Sana}, H., \& {Marchant}, P. 2023{\natexlab{b}},
  \aap, 670, A79, \dodoi{10.1051/0004-6361/202244818}

\bibitem[{{Kaplan} \& {Makarov}(2003)}]{2003AN....324..419K}
{Kaplan}, G.~H., \& {Makarov}, V.~V. 2003, Astronomische Nachrichten, 324, 419,
  \dodoi{10.1002/asna.200310159}

\bibitem[{{Kervella} {et~al.}(2019){Kervella}, {Arenou}, {Mignard}, \&
  {Th{\'e}venin}}]{2019A&A...623A..72K}
{Kervella}, P., {Arenou}, F., {Mignard}, F., \& {Th{\'e}venin}, F. 2019, \aap,
  623, A72, \dodoi{10.1051/0004-6361/201834371}

\bibitem[{{Kervella} {et~al.}(2022){Kervella}, {Arenou}, \&
  {Th{\'e}venin}}]{2022A&A...657A...7K}
{Kervella}, P., {Arenou}, F., \& {Th{\'e}venin}, F. 2022, \aap, 657, A7,
  \dodoi{10.1051/0004-6361/202142146}

\bibitem[{{Lindegren} {et~al.}(2021){Lindegren}, {Klioner}, {Hern{\'a}ndez},
  {Bombrun}, {Ramos-Lerate}, {Steidelm{\"u}ller}, {Bastian}, {Biermann}, {de
  Torres}, {Gerlach}, {Geyer}, {Hilger}, {Hobbs}, {Lammers}, {McMillan},
  {Stephenson}, {Casta{\~n}eda}, {Davidson}, {Fabricius}, {Gracia-Abril},
  {Portell}, {Rowell}, {Teyssier}, {Torra}, {Bartolom{\'e}}, {Clotet},
  {Garralda}, {Gonz{\'a}lez-Vidal}, {Torra}, {Abbas}, {Altmann}, {Anglada
  Varela}, {Balaguer-N{\'u}{\~n}ez}, {Balog}, {Barache}, {Becciani}, {Bernet},
  {Bertone}, {Bianchi}, {Bouquillon}, {Brown}, {Bucciarelli}, {Busonero},
  {Butkevich}, {Buzzi}, {Cancelliere}, {Carlucci}, {Charlot}, {Cioni},
  {Crosta}, {Crowley}, {del Peloso}, {del Pozo}, {Drimmel}, {Esquej}, {Fienga},
  {Fraile}, {Gai}, {Garcia-Reinaldos}, {Guerra}, {Hambly}, {Hauser},
  {Jan{\ss}en}, {Jordan}, {Kostrzewa-Rutkowska}, {Lattanzi}, {Liao}, {Licata},
  {Lister}, {L{\"o}ffler}, {Marchant}, {Masip}, {Mignard}, {Mints}, {Molina},
  {Mora}, {Morbidelli}, {Murphy}, {Pagani}, {Panuzzo}, {Pe{\~n}alosa Esteller},
  {Poggio}, {Re Fiorentin}, {Riva}, {Sagrist{\`a} Sell{\'e}s}, {Sanchez
  Gimenez}, {Sarasso}, {Sciacca}, {Siddiqui}, {Smart}, {Souami}, {Spagna},
  {Steele}, {Taris}, {Utrilla}, {van Reeven}, \&
  {Vecchiato}}]{2021A&A...649A...2L}
{Lindegren}, L., {Klioner}, S.~A., {Hern{\'a}ndez}, J., {et~al.} 2021, \aap,
  649, A2, \dodoi{10.1051/0004-6361/202039709}

\bibitem[{{Lockwood} \& {Wing}(1982)}]{1982MNRAS.198..385L}
{Lockwood}, G.~W., \& {Wing}, R.~F. 1982, \mnras, 198, 385,
  \dodoi{10.1093/mnras/198.2.385}

\bibitem[{{Mahy} {et~al.}(2022){Mahy}, {Sana}, {Shenar}, {Sen}, {Langer},
  {Marchant}, {Abdul-Masih}, {Banyard}, {Bodensteiner}, {Bowman}, {Dsilva},
  {Fabry}, {Hawcroft}, {Janssens}, {Van Reeth}, \& {Eldridge}}]{Mahy2022}
{Mahy}, L., {Sana}, H., {Shenar}, T., {et~al.} 2022, \aap, 664, A159,
  \dodoi{10.1051/0004-6361/202243147}

\bibitem[{{Makarov}(2022)}]{2022AJ....164..157M}
{Makarov}, V.~V. 2022, \aj, 164, 157, \dodoi{10.3847/1538-3881/ac88d1}

\bibitem[{{Makarov} \& {Kaplan}(2005)}]{2005AJ....129.2420M}
{Makarov}, V.~V., \& {Kaplan}, G.~H. 2005, \aj, 129, 2420,
  \dodoi{10.1086/429590}

\bibitem[{{Makarov} \& {Murphy}(2007)}]{2007AJ....134..367M}
{Makarov}, V.~V., \& {Murphy}, D.~W. 2007, \aj, 134, 367,
  \dodoi{10.1086/518242}

\bibitem[{{Makarov} \& {Tokovinin}(2019)}]{2019AJ....157..136M}
{Makarov}, V.~V., \& {Tokovinin}, A. 2019, \aj, 157, 136,
  \dodoi{10.3847/1538-3881/ab05e0}

\bibitem[{{Margoni} {et~al.}(1988){Margoni}, {Stagni}, \&
  {Mammano}}]{1988A&AS...75..157M}
{Margoni}, R., {Stagni}, R., \& {Mammano}, A. 1988, \aaps, 75, 157

\bibitem[{{Mignard} \& {Klioner}(2012)}]{2012A&A...547A..59M}
{Mignard}, F., \& {Klioner}, S. 2012, \aap, 547, A59,
  \dodoi{10.1051/0004-6361/201219927}

\bibitem[{{Mikaelian}(1978)}]{1978Ap&SS..57..245M}
{Mikaelian}, K.~O. 1978, \apss, 57, 245, \dodoi{10.1007/BF00639061}

\bibitem[{{Murray}(1983)}]{1983veas.book.....M}
{Murray}, C.~A. 1983, {Vectorial astrometry}

\bibitem[{{Palacios} {et~al.}(2010){Palacios}, {Gebran}, {Josselin}, {Martins},
  {Plez}, {Belmas}, \& {L{\`e}bre}}]{2010A&A...516A..13P}
{Palacios}, A., {Gebran}, M., {Josselin}, E., {et~al.} 2010, \aap, 516, A13,
  \dodoi{10.1051/0004-6361/200913932}

\bibitem[{{Perryman} {et~al.}(1997){Perryman}, {Lindegren}, {Kovalevsky},
  {Hoeg}, {Bastian}, {Bernacca}, {Cr{\'e}z{\'e}}, {Donati}, {Grenon},
  {Grewing}, {van Leeuwen}, {van der Marel}, {Mignard}, {Murray}, {Le Poole},
  {Schrijver}, {Turon}, {Arenou}, {Froeschl{\'e}}, \&
  {Petersen}}]{1997A&A...323L..49P}
{Perryman}, M.~A.~C., {Lindegren}, L., {Kovalevsky}, J., {et~al.} 1997, \aap,
  323, L49

\bibitem[{{Przybylski}(1972)}]{1972MNRAS.159..155P}
{Przybylski}, A. 1972, \mnras, 159, 155, \dodoi{10.1093/mnras/159.2.155}

\bibitem[{{Tabernero} {et~al.}(2021){Tabernero}, {Dorda}, {Negueruela}, \&
  {Marfil}}]{2021A&A...646A..98T}
{Tabernero}, H.~M., {Dorda}, R., {Negueruela}, I., \& {Marfil}, E. 2021, \aap,
  646, A98, \dodoi{10.1051/0004-6361/202039236}

\bibitem[{{Titov} \& {Malkin}(2009)}]{2009A&A...506.1477T}
{Titov}, O., \& {Malkin}, Z. 2009, \aap, 506, 1477,
  \dodoi{10.1051/0004-6361/200912369}

\bibitem[{{Tokovinin}(2016)}]{chiron1}
{Tokovinin}, A. 2016, \aj, 152, 11, \dodoi{10.3847/0004-6256/152/1/11}

\bibitem[{{Tokovinin}(2018)}]{2018ApJS..235....6T}
---. 2018, \apjs, 235, 6, \dodoi{10.3847/1538-4365/aaa1a5}

\bibitem[{{Tokovinin} {et~al.}(2013){Tokovinin}, {Fischer}, {Bonati},
  {Giguere}, {Moore}, {Schwab}, {Spronck}, \& {Szymkowiak}}]{CHIRON}
{Tokovinin}, A., {Fischer}, D.~A., {Bonati}, M., {et~al.} 2013, \pasp, 125,
  1336, \dodoi{10.1086/674012}

\bibitem[{{Tokovinin} {et~al.}(2022){Tokovinin}, {Mason}, {Mendez}, \&
  {Costa}}]{2022AJ....164...58T}
{Tokovinin}, A., {Mason}, B.~D., {Mendez}, R.~A., \& {Costa}, E. 2022, \aj,
  164, 58, \dodoi{10.3847/1538-3881/ac78e7}

\bibitem[{{Tokovinin} {et~al.}(2020){Tokovinin}, {Mason}, {Mendez}, {Costa}, \&
  {Horch}}]{2020AJ....160....7T}
{Tokovinin}, A., {Mason}, B.~D., {Mendez}, R.~A., {Costa}, E., \& {Horch},
  E.~P. 2020, \aj, 160, 7, \dodoi{10.3847/1538-3881/ab91c1}

\bibitem[{{Vityazev} \& {Shuksto}(2004)}]{2004ASPC..316..230V}
{Vityazev}, V., \& {Shuksto}, A. 2004, in Astronomical Society of the Pacific
  Conference Series, Vol. 316, Order and Chaos in Stellar and Planetary
  Systems, ed. G.~G. {Byrd}, K.~V. {Kholshevnikov}, A.~A. {Myllri}, I.~I.
  {Nikiforov}, \& V.~V. {Orlov}, 230

\bibitem[{{Vityazev}(2010)}]{2010AIPC.1283...94V}
{Vityazev}, V.~V. 2010, in American Institute of Physics Conference Series,
  Vol. 1283, Mathematics and Astronomy: A Joint Long Journey, ed. M.~{de
  Le{\'o}n}, D.~M. {de Diego}, \& R.~M. {Ros} (AIP), 94--113,
  \dodoi{10.1063/1.3506085}

\bibitem[{{Whiting} {et~al.}(2023){Whiting}, {Hill}, {Bromley}, \&
  {Kenyon}}]{2023AJ....165..193W}
{Whiting}, M.~L., {Hill}, J.~B., {Bromley}, B.~C., \& {Kenyon}, S.~J. 2023,
  \aj, 165, 193, \dodoi{10.3847/1538-3881/acc526}

\bibitem[{{Woosley}(1993)}]{1993ApJ...405..273W}
{Woosley}, S.~E. 1993, \apj, 405, 273, \dodoi{10.1086/172359}

\bibitem[{{Xu} {et~al.}(2018){Xu}, {Zhang}, {Reid}, {Menten}, {Zheng}, \&
  {Wang}}]{2018ApJ...859...14X}
{Xu}, S., {Zhang}, B., {Reid}, M.~J., {et~al.} 2018, \apj, 859, 14,
  \dodoi{10.3847/1538-4357/aabba6}

\bibitem[{{Zak} {et~al.}(2023){Zak}, {Jones}, {Boffin}, {Beck}, {Klencki},
  {Bodensteiner}, {Shenar}, {Van Winckel}, {Skarka}, {Arellano-C{\'o}rdova},
  {Viuho}, {Sowicka}, {Guenther}, \& {Hatzes}}]{Zak2023}
{Zak}, J., {Jones}, D., {Boffin}, H.~M.~J., {et~al.} 2023, \mnras, 524, 5749,
  \dodoi{10.1093/mnras/stad2137}

\bibitem[{{Zeldovich} \& {Guseynov}(1966)}]{1966ApJ...144..840Z}
{Zeldovich}, Y.~B., \& {Guseynov}, O.~H. 1966, \apj, 144, 840,
  \dodoi{10.1086/148672}

\end{thebibliography}
\bibliographystyle{aasjournal}

%\appendix
%\section{Objects of note and peculiar sources}

\end{document}